\documentclass[nonacm,sigplan,authorversion]{acmart}

\usepackage[]{hyperref}

\usepackage{tikz}
\usepackage[english]{babel}
\usepackage{blindtext}

\usepackage{breakurl}
\usepackage{tightenum}
\usepackage{wrapfig}
\usepackage{pifont}

\usepackage{enumitem}
\usepackage{listings}
\usepackage{latexsym}
\usepackage{adjustbox}
\usepackage{graphicx}
\usepackage{floatflt}
\usepackage{setspace}
\usepackage{graphicx}
\usepackage{algorithm}
\usepackage{amsmath}

\usepackage{xspace}
\usepackage{makecell}
\usepackage{caption}

\usepackage[normalem]{ulem}

\usepackage{svg}
\usepackage{subfigure}

\newcommand{\ke}[1]{{\color{red}Ke:  {#1}}}
\newcommand{\xu}[1]{{{#1}}}

\newcommand{\theonote}[1]{}

\newcommand{\nonapproximate}[1]{non-approximate\xspace}

\usepackage{calc}

\usepackage[textsize=tiny,textwidth=0.6in]{todonotes}
\newcommand{\allnotes}[1]{}
\newcommand{\ignore}[1]{}

\newcommand{\boldparagraph}[1]{\vspace*{1ex}\noindent\textbf{#1}\hspace{1em}}

\def\cfigure[#1,#2,#3]{
\begin{figure}
\vspace*{0mm}
\begin{center}

\includegraphics[width=3in]{#1} 
 
\vspace*{-3mm}\caption[]{#2
} \label{#3}
 
\vspace*{-5mm}
\end{center}
\end{figure}}

\def\cfigurefour[#1,#2,#3]{
\begin{figure}
\vspace*{0mm}
\begin{center}

\includegraphics[width=4in]{#1} 
 
\vspace*{-3mm}\caption[]{#2
} \label{#3}
 
\vspace*{-5mm}
\end{center}
\end{figure}}

\def\cfiguretemp[#1,#2,#3]{
\begin{figure}
\vspace*{0mm}
\begin{center}

\includegraphics[width=3.5in]{#1} 
 
\vspace*{-3mm}\caption[]{#2
} \label{#3}
 
\vspace*{-5mm}
\end{center}
\vspace*{-2mm}
\end{figure}}

\def\wfigure[#1,#2,#3]{
\begin{figure*}
\vspace*{0mm}
\begin{center}
 \includegraphics[width=\textwidth]{#1} 
 \vspace*{-3mm}\caption[]{#2
} \label{#3}
 
\end{center}
\end{figure*}}

\def\threefigure[#1,#2,#3,#4,#5]{
\begin{figure*}
\vspace*{0mm}
\begin{center}

\begin{tabular}{ccc}
\includegraphics[width=2in]{#1} & \includegraphics[width=2in]{#2} &  \includegraphics[width=2in]{#3} \\
(a) & (b) & (c) \\
\end{tabular}

\vspace*{-3mm}\caption[]{#4
} \label{#5}

\vspace*{-5mm}
\end{center}
\vspace*{-2mm}
\end{figure*}}

\def\dcfigure[#1,#2,#3,#4,#5,#6]{
{
\begin{figure*}
\begin{center}
\begin{minipage}[c]{\columnwidth}{
\includegraphics[width=\columnwidth]{#1} 
\vspace*{0mm}\caption[]{#2} \label{#3} \
}\end{minipage}\hspace*{\columnsep}\
\begin{minipage}[c]{\columnwidth}{
\includegraphics[width=\columnwidth]{#4} 
\vspace*{0mm}\caption[]{#5}\label{#6} \
}\end{minipage}
\end{center}
\end{figure*}
}
}

\def\tableByTable[#1,#2,#3,#4,#5,#6]{
{
\begin{table*}
\begin{center}
\begin{minipage}[c]{3in}{
\centering
{#1}
\vspace*{0mm}\tabcaption[]{#2}\label{#3} \
}\end{minipage}\hspace*{\columnsep}\
\begin{minipage}[c]{3in}{
\centering
{#4}
\vspace*{0mm}\tabcaption[]{#5}\label{#6} \
}\end{minipage}
\end{center}
\end{table*}
}
}

\def\figureByTable[#1,#2,#3,#4,#5,#6]{
{
\begin{figure*}
\begin{center}
\begin{minipage}[c]{3in}{
\centering
\includegraphics[width=\textwidth]{#1}
\vspace*{0mm}\figcaption[]{#2} \label{#3} \
}\end{minipage}\hspace*{\columnsep}\
\begin{minipage}[c]{3.3in}{
\centering
{#4}
\vspace*{0mm}\tabcaption[]{#5}\label{#6} \
}\end{minipage}
\end{center}
\end{figure*}
}
}

\def\tableByFigure[#1,#2,#3,#4,#5,#6]{
{
\begin{figure*}
\begin{center}
\begin{minipage}[c]{4.3in}{
\centering
{#1}
\vspace*{0mm}\tabcaption[]{#2} \label{#3} \
}\end{minipage}\hspace*{\columnsep}\
\begin{minipage}[c]{2.2in}{
\centering
\includegraphics[width=\textwidth]{#4}
\vspace*{-0.35in}\caption[]{#5}\label{#6} \
}\end{minipage}
\end{center}
\end{figure*}
}
}

\def\doublecfigure[#1,#2,#3,#4]{
{
\begin{figure}
\begin{center}
\begin{minipage}[c]{1.5in}{
\begin{center}
\includegraphics[width=1.5in]{#1}
\end{center}
}\end{minipage}\hspace*{1em}\
\begin{minipage}[c]{1.5in}{
\begin{center}
\includegraphics[width=1.5in]{#2}
\end{center}
}\end{minipage}
\vspace*{0mm}\caption[]{#3} \label{#4} \
\end{center}
\end{figure}
}
}

\def\qcfigure[#1,#2,#3,#4,#5,#6]{
{
\begin{figure*}
\vspace*{0.2in}\
\begin{center}
\begin{minipage}[c]{3in}{
\includegraphics[width=3in]{#1} 
\vspace*{-3mm}
}
\end{minipage}\hspace*{0.5in}\
\begin{minipage}[c]{3in}{
\includegraphics[width=3in]{#2} 
\vspace*{-3mm}
}\end{minipage}

\begin{minipage}[c]{3in}{
\includegraphics[width=3in]{#3} 
\vspace*{-3mm}
}
\end{minipage}\hspace*{0.5in}\
\begin{minipage}[c]{3in}{
\includegraphics[width=3in]{#4} 
\vspace*{-3mm}
}\end{minipage}
\end{center}
\caption[]{#5}\label{#6}
\end{figure*}
}
}

\def\twfigure[#1,#2,#3,#4,#5]{
{
\begin{figure*}
\vspace*{0.2in}\
\begin{center}
\begin{minipage}[c]{6.5in}{
\includegraphics[width=6.5in]{#1} 
\vspace*{-3mm}
}
\end{minipage}

\begin{minipage}[c]{6.5in}{
\includegraphics[width=6.5in]{#2} 
\vspace*{-3mm}
}\end{minipage}

\begin{minipage}[c]{6.5in}{
\includegraphics[width=6.5in]{#3} 
\vspace*{-3mm}
}
\end{minipage}
\end{center}
\caption[]{#4}\label{#5}
\end{figure*}
}
}

\def\dwfigure[#1,#2,#3,#4]{
{
\begin{figure*}
\vspace*{0.2in}\
\begin{center}
\begin{minipage}[c]{6.5in}{
\includegraphics[width=6.5in]{#1} 
\vspace*{-3mm}
}
\end{minipage}

\begin{minipage}[c]{6.5in}{
\includegraphics[width=6.5in]{#2} 
\vspace*{-3mm}
}\end{minipage}

\end{center}
\caption[]{#3}\label{#4}
\end{figure*}
}
}

\def\dssfigure[#1,#2,#3,#4,#5,#6]{
{
\begin{figure*}
\vspace*{0.2in}\
\begin{center}
\begin{minipage}[c]{4in}{
\includegraphics[width=4in]{#1}
\vspace*{-3mm}\caption[]{#2} \label{#3} \
}\end{minipage}\hspace*{0.5in}\
\begin{minipage}[c]{2in}{
\includegraphics[width=2in]{#4}
\vspace*{-3mm}\caption[]{#5}\label{#6} \
}\end{minipage}
\end{center}
\vspace*{-0.4in}\
\end{figure*}
}
}

\def\dsfigure[#1,#2,#3,#4,#5,#6]{
{
\begin{figure*}
\vspace*{0.2in}\
\begin{center}
\begin{minipage}[c]{3in}{
\includegraphics[width=3in]{#1}
\vspace*{-3mm}\caption[]{#2} \label{#3} \
}\end{minipage}\hspace*{0.5in}\
\begin{minipage}[c]{3in}{
\hspace*{0.5in}\
\includegraphics[height=3in]{#4}
\vspace*{-3mm}\caption[]{#5}\label{#6} \
}\end{minipage}
\end{center}
\vspace*{-0.4in}\
\end{figure*}
}
}

\def\dsyfigure[#1,#2,#3,#4,#5,#6]{
{
\begin{figure*}
\vspace*{0.2in}\
\begin{center}
\begin{minipage}[c]{2.5in}{
\includegraphics[height=2.5in]{#1}
\vspace*{-3mm}\caption[]{#2} \label{#3} \
}\end{minipage}\hspace*{0.5in}\
\begin{minipage}[c]{2.5in}{
\includegraphics[height=2.5in]{#4}
\vspace*{-3mm}\caption[]{#5}\label{#6} \
}\end{minipage}
\end{center}
\vspace*{-0.4in}\
\end{figure*}
}
}

\def\dyfigure[#1,#2,#3,#4,#5,#6]{
{
\begin{figure*}
\vspace*{0.2in}\
\begin{center}
\begin{minipage}[c]{3in}{
\includegraphics[height=3in]{#1} 
\vspace*{-3mm}\caption[]{#2} \label{#3} \
}\end{minipage}\hspace*{0.5in}\
\begin{minipage}[c]{3in}{
\includegraphics[height=3in]{#4} 
\vspace*{-3mm}\caption[]{#5}\label{#6} \
}\end{minipage}
\end{center}
\vspace*{-0.4in}\
\end{figure*}
}
}

\def\dyoldfigure[#1,#2,#3,#4,#5,#6]{
{
\begin{figure*}
\vspace*{0.2in}\
\begin{center}
\begin{minipage}[c]{3in}{
\epsfysize=2.0in\
\hspace{0.5in}\
\epsfbox{#1}
\vspace*{-3mm}\caption[]{#2} \label{#3} \
}\end{minipage}\hspace*{0.25in}\
\begin{minipage}[c]{3in}{
\epsfysize=2.0in\
\hspace{0.5in}\
\epsfbox{#4}
\vspace*{-3mm}\caption[]{#5}\label{#6} \
}\end{minipage}
\end{center}
\vspace*{-0.4in}\
\end{figure*}
}
}

\def\cfiguredouble[#1,#2,#3,#4]{
\begin{figure}
\vspace*{0.2in}\
\begin{center}
\begin{minipage}[c]{1.5in}{
\epsfxsize=1.5in\
\epsfbox{#1}
}\end{minipage}\hspace*{0.1in}\
\begin{minipage}[c]{1.5in}{
\epsfxsize=1.5in\
\vspace{0.1in}\epsfbox{#2}
}\end{minipage}\vspace*{-0.10in} \caption[]{#3}\label{#4}
\end{center}
\vspace*{-0.4in}\
\end{figure}
}

\def\wpfigure[#1,#2,#3,#4]{
\begin{figure*}
\vspace*{4mm}
\begin{center}

\includegraphics[width=#4]{#1} 

\vspace*{-3mm}\caption[]{#2
} \label{#3}

\vspace*{-5mm}
\end{center}
\end{figure*}}

\def\wprfigure[#1,#2,#3,#4,#5]{
\begin{figure*}
\vspace*{4mm}
\begin{center}

\includegraphics[width=#4, angle=#5]{#1} 

\vspace*{-3mm}\caption[]{#2
} \label{#3}

\vspace*{-5mm}
\end{center}
\end{figure*}}

\def\DoubleFigureWSlide[#1,#2,#3,#4,#5,#6,#7,#8,#9]{
\begin{figure*}
\vspace*{#9}
\begin{center}
\begin{minipage}{#4}
\includegraphics[width=#4]{#1}
\vspace*{-3mm}\caption{#2
}\label{#3}
\end{minipage}
\hspace{2em}
\begin{minipage}{#8}
\includegraphics[width=#8]{#5}
\vspace*{-3mm}\caption{#6
}\label{#7}
\end{minipage}
\vspace*{-5mm}
\end{center}
\end{figure*}
}

\def\DoubleFigureW[#1,#2,#3,#4,#5,#6,#7,#8]{
\begin{figure*}
\vspace*{0in}
\begin{center}
\begin{minipage}{#4}
\includegraphics[width=#4]{#1}
\vspace*{-3mm}\caption{#2
}\label{#3}
\end{minipage}
\hspace{2em}
\begin{minipage}{#8}
\includegraphics[width=#8]{#5}
\vspace*{-3mm}\caption{#6
}\label{#7}
\end{minipage}
\vspace*{-5mm}
\end{center}
\end{figure*}
}

\def\DoubleFigureWHack[#1,#2,#3,#4,#5,#6,#7,#8]{
\begin{figure*}
\vspace*{0in}
\begin{center}
\begin{minipage}{3in}
\includegraphics[width=#4]{#1}
\vspace*{-3mm}\caption{#2
}\label{#3}
\end{minipage}
\hspace{2em}
\begin{minipage}{3in}
\includegraphics[width=#8]{#5}
\vspace*{-3mm}\caption{#6
}\label{#7}
\end{minipage}
\vspace*{-5mm}
\end{center}
\end{figure*}
}

\def\ddcfigure[#1,#2,#3,#4]{
\begin{figure*}
\vspace*{0.2in}\
\begin{center}
\begin{minipage}[c]{\columnwidth}{
\includegraphics[width=\columnwidth]{#1} 
}\end{minipage}\hspace{0.5in}\
\begin{minipage}[c]{\columnwidth}{
\includegraphics[width=\columnwidth]{#2} 
}\end{minipage} \caption[]{#3}\label{#4}
\end{center}
\end{figure*}
}

\def\ddcfigureSlide[#1,#2,#3,#4,#5]{
\begin{figure*}
\vspace*{#5}\
\begin{center}
\begin{minipage}[c]{3in}{
\includegraphics[height=3in]{#1} 
}\end{minipage}\hspace{0.5in}\
\begin{minipage}[c]{3in}{
\includegraphics[height=3in]{#2} 
}\end{minipage}\vspace*{-0.10in} \caption[]{#3}\label{#4}
\end{center}
\vspace*{-0.4in}\
\end{figure*}
}

\def\cxfigure[#1,#2,#3]{
\begin{figure}
\vspace*{4mm}
\begin{center}
 
\epsfxsize=2.5in\
\epsfbox{#1}\
 
\vspace*{-0.10in}\caption[]{#2
} \label{#3}
 
\vspace*{-5mm}
\end{center}
\vspace*{-2mm}
\end{figure}}

\newcommand{\beforecaption}{\vspace{-.15cm}\begin{spacing}{0.85}}
\newcommand{\aftercaption}{\vspace{-.45cm}\end{spacing}}

\newcommand{\eg}{\textit{e.g.}}
\newcommand{\ie}{\textit{i.e.}}

\newcommand{\etc}{\textit{etc.}}

\newcommand{\KB}{\,KB}
\newcommand{\MB}{\,MB}

\newcommand{\gbps}{\,Gbps}

\newcommand{\sysname}{DFabric\xspace}

\newcommand{\FBKV}[1]{FBKV\xspace}
\newcommand{\FBHD}[1]{FBHD\xspace}

\newcommand{\cxl}{CXL\xspace}
\newcommand{\load}{load\xspace}
\newcommand{\store}{store\xspace}
\newcommand{\cxlmem}{CXL.mem\xspace}
\newcommand{\cxlcache}{CXL.cache\xspace}
\newcommand{\cxlio}{CXL.io\xspace}
\newcommand{\dma}{DMA\xspace}
\newcommand{\send}{\texttt{SEND}\xspace}
\newcommand{\recv}{\texttt{RECV}\xspace}
\newcommand{\dcn}{DCN\xspace}
\newcommand{\dnn}{DNN\xspace}
\newcommand{\dra}{DRA\xspace}
\newcommand{\pcie}{PCIe\xspace}
\newcommand{\nic}{NIC\xspace}
\newcommand{\tor}{ToR\xspace}

\newcommand{\ntb}{NTB\xspace}

\newcommand{\bsp}{BSP\xspace}
\newcommand{\cn}{CN\xspace}
\newcommand{\fabric}{CXL fabric\xspace}
\newcommand{\lppu}{LPPU\xspace}
\newcommand{\vnic}{vNIC\xspace}

\newcommand{\sect}{\texttt{Section}\xspace}
\newcommand{\buffer}{\texttt{Buffer}\xspace}
\newcommand{\region}{\texttt{Region}\xspace}
\newcommand{\vq}{\texttt{virt\_queue}\xspace}
\newcommand{\rxq}{\texttt{RxQ}\xspace}
\newcommand{\txq}{\texttt{TxQ}\xspace}
\newcommand{\opq}{\texttt{OpQ}\xspace}
\newcommand{\phyq}{\texttt{physical\_queue}\xspace}
\newcommand{\cq}{\texttt{completion\_phq}\xspace}
\newcommand{\asic}{ASIC\xspace}
\newcommand{\sn}{SN\xspace}

\newcommand{\doce}{DoCE\xspace}

\newcommand{\nvlink}{NVLink\xspace}
\newcommand{\gpu}{GPU\xspace}

\newcommand{\allreduce}{Allreduce\xspace}

\usepackage{xcolor}
\definecolor{commentgreen}{RGB}{2,112,10}
\definecolor{eminence}{RGB}{108,48,130}
\definecolor{weborange}{RGB}{255,165,0}
\definecolor{frenchplum}{RGB}{129,20,83}

\begin{document}



\title{\sysname: Scaling Out Data Parallel Applications with CXL-Ethernet Hybrid Interconnects}







\author{Xu Zhang}
\affiliation{%
  \country{China}
}
\email{zhangxu19s@ict.ac.cn}

\author{Ke Liu}
\affiliation{%
  \country{China}
}
\email{liuke@ict.ac.cn}

\author{Yisong Chang}
\affiliation{%
  \country{China}
}
\email{changyisong@ict.ac.cn}



\author{Ke Zhang}
\affiliation{%
  \country{China}
}
\email{zhangke@ict.ac.cn}

\author{Mingyu Chen}
\affiliation{%
  \country{China}
}
\email{cmy@ict.ac.cn}

\begin{abstract}
%

Emerging interconnects, such as \cxl and \nvlink, 
have been integrated into the intra-host topology to scale more accelerators and 
facilitate efficient communication between them, such as \gpu{}s.
To keep pace with the accelerator's growing computing throughput, 
the interconnect has seen substantial enhancement in link bandwidth, 
\eg, 256GBps for \cxl 3.0 links, 
which surpasses Ethernet and InfiniBand network links by an order of magnitude or more.
Consequently, 
when data-intensive jobs, such as LLM training, 
scale across multiple hosts beyond the reach limit of the interconnect,
the performance is significantly hindered by the limiting bandwidth of the network infrastructure.
%
We address the problem by proposing \sysname, a two-tier interconnect architecture.
First, 
\sysname disaggregates rack's computing units with an interconnect fabric, \ie, \cxl fabric, which scales at rack-level, so that they can enjoy intra-rack efficient interconnecting.
Second,
\sysname disaggregates \nic{}s from hosts, 
and consolidates them to form a \nic pool with \cxl fabric.
By providing sufficient aggregated capacity comparable to interconnect bandwidth,
the \nic pool bridges efficient communication across racks or beyond the reach limit of interconnect fabric.
However, 
the local memory accessing becomes the bottleneck when enabling each host to efficiently utilize the \nic pool.
To the end, \sysname builds a memory pool with sufficient bandwidth by disaggregating host local memory and adding more memory devices. 
%
We have implemented a prototype of \sysname that can run applications transparently. 
We validated its performance gain by running various microbenchmarks and compute-intensive applications such as DNN and graph.

\end{abstract}

\maketitle
\thispagestyle{plain}

\section{Introduction}
\label{sec:intro}

Data-intensive applications running in today's production clouds, such as graph processing~\cite{SIGMOD10Pregel}, data analytics~\cite{Commun08mapreduce}, and deep neural network (\dnn) training~\cite{Neurocomputing17DNN} often operate with Bulk Synchronous Parallel (\bsp)~\cite{Commun90BSP} or MapReduce paradigms,
which requires iterative execution of two key stages:
data computation across multiple computing units 
(\eg, forward and backward propagation in \dnn training), 
and communication between these units (\eg, data shuffling in MapReduce).
Such applications are compute-intensive and would be very time-consuming.
For example, 
training a BERT model on a single TPU takes over 1.5 months~\cite{arXiv18bert}. 

To accelerate the process, 
emerging interconnects, such as \cxl (Compute Express Link)~\cite{CXLSpec} and \nvlink~\cite{Micro17NVLink}, 
have been integrated into the intra-host interconnect topology to scale a host with an increased number of computing nodes (\eg, accelerator, \gpu{}, TPU~\cite{ISCA23tpu}) 
and to facilitate efficient communication between them, for example, by supporting high bandwidth direct point-to-point communication.
To keep pace with the evolving computing throughput, 
the interconnect has experienced a substantial improvement in link bandwidth
as shown in Table~\ref{tab:bandwidth}.
The link bandwidth of those new interconnects exceeds that of conventional network links 
by an order of magnitude or more, and this gap is expected to expand rapidly~\cite{SIGMOD20pump}.

\begin{table}
    \centering
    \caption{The latency and bandwidth of mainstream devices.}
    \small
    \begin{tabular}{|c|c|c|c|c|c|}
    \hline
         &\multicolumn{2}{|c|}{Memory}&  \multicolumn{2}{|c|}{Interconnect}&  Network\\
         \hline
         Device &DDR5&  GDDR6 &  \makecell[c]{CXL \\3.0}&  \makecell[c]{NVLink \\4.0}& \makecell[c]{InfiniBand\\/ Ethernet}\\
         \hline
         Latency & <80ns & <220ns & \makecell[c]{<400ns\\\cite{MICRO23sun}}& <8us~\cite{TPDS20Li} & >2us~\cite{EuroSys23DiLOS} \\
         \hline
         B.W.& 50 & 400 & 120 & 900 & 25/50 \\
         \hline
    \end{tabular}
    \label{tab:bandwidth}
    \scriptsize{
$\ast$The bandwidth is measured in GBps.
}
\end{table}

However, 
due to the distance limit of interconnect bus 
(existing \cxl is up to 2-m maximum distance~\cite{SC22CXL3}), 
they have limited scale, \eg, rack level~\cite{TACO24Rcmp}.
Thereby, the common practice in enterprise or public clouds is with emerging fast interconnects (\eg, \nvlink, \cxl) within hosts and network interconnects (\eg, Ethernet) between hosts~\cite{arXiv20scalable}.
Due to the huge gap between interconnect bandwidth and network bandwidth, 
as will show in \S\ref{sec:motivation},
the data-parallel job, such as \allreduce, 
across multiple hosts or racks beyond the reach limit of the interconnect,
its communication efficiency is inevitably hindered by the slow network link between hosts. 

The intuitive approach to addressing the gap is to build a \nic pool with its aggregated capacity larger than that of the interconnect, 
intended for efficient communications beyond the interconnect's reach limit.
However, 
building a separate pool for each node's local interconnect is cost-prohibitive due to the significant bandwidth disparity. For example, to match the bandwidth of a \cxl 3.0 link, it would be necessary to install more than ten $200$\gbps\ \nic{}s per host.
Instead, 
we propose connecting hosts and computing nodes using fast interconnects at a larger scale, 
\ie, rack level, 
thus there are sufficient number of existing \nic{}s from hosts to be leveraged  
to form a \nic pool for communications across racks.

\begin{figure}[t]
\centering
  \includegraphics[width=0.45\textwidth]{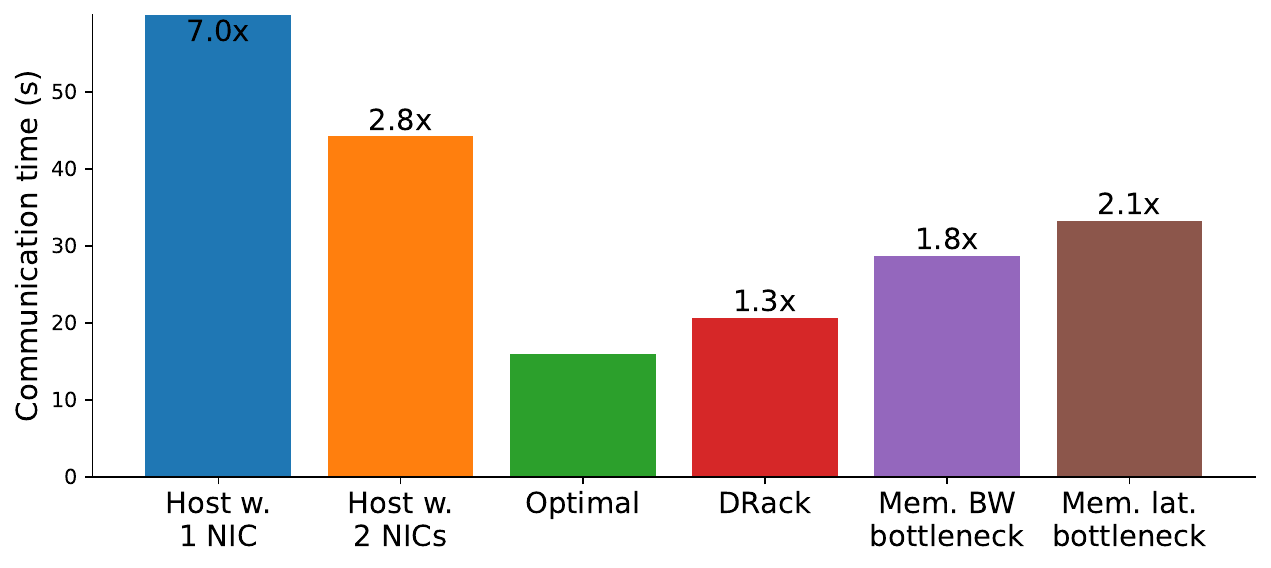}

\caption{
The communication time under different bottlenecks running the ring allreduce.
}
  \label{fig:motivation}
\end{figure}

Specifically, 
we propose \sysname, 
a two-tier interconnect architecture. 
For intra-rack interconnect, 
\sysname utilizes \cxl fabric, 
a recent interconnect fabric introduced in \cxl 3.0, 
capable of scaling to thousands of different types of nodes like processors, memory devices, accelerators, and \nic{}s.
The fabric extends intra-host interconnects to the rack level.
Communications across any pair of computing nodes (\cn) across hosts 
is ensured by the fabric's high bi-sectional bandwidth.
For inter-rack level interconnect,
\sysname also disaggregates existing \nic{}s from individual hosts using the CXL fabric, and aggregates them into a \nic pool.
The \nic pool is accessed by any node within the rack and tends to offer a sufficient aggregate bandwidth to transfer across-rack traffic load.
Given the insight that \nic{}s in conventional datacenter racks are often underutilized 
due to the on-off traffic pattern, as measured in ~\cite{NSDI15GRIN}, 
and the contention at the \nic pool can be minimized by using some data parallelism approaches that exploit traffic locality, \eg, hierarchical \allreduce, 
or adopting contention-free communication patterns, \eg, ring \allreduce, by expanding the pool capacity with additional \nic{}s,  
\sysname achieves the goal
of providing a cost-effective, efficient, and flexible means to bridge the bandwidth gap in inter-rack communications.
We will show in Figure~\ref{fig:motivation} that \sysname achieves the optimal communication efficiency by simply forming a \nic pool with existing \nic{}s, when running ring \allreduce, where only one node is communicating with the pool.

However, 
the memory bandwidth (\xu{The large NIC pool may push bottleneck to the Integrated IO controller~\cite{Anuj16ATC, PicNIC19SIGCOMM, Rolf18SIGCOMM, Midhul24SIGCOMM, Hostping23NSDI} and limited number of memory channels located in each host.}) may pose a bottleneck that hinders the full utilization of the \nic pool's capacity. 
This is because the bandwidth of local memory, \ie, system-integrated memory, 
or a specific memory device like GDDR, can be less than the capacity of the \nic pool. 
Consequently, when performing direct memory access (\dma) from all \nic{}s, the memory bandwidth becomes the constraining factor, reducing the achievable throughput of the \nic pool.
%
%
As \cxl fabric in \cxl 3.0~\cite{CXLSpec} allows devices to participate in host processor’s coherence domain while accessing memory, 
we disaggregate the local memory and additional memory devices a rack using \cxl fabric, 
and map them into a single memory address space to build a shared memory pool that can be accessed coherently by any device. 
Thus, 
the network traffic received from the \nic pool can be written into multiple memory devices with the aggregated memory bandwidth larger than the \nic pool's capacity. 
Similarly, \nic pool reads the data using \dma from multiple memory devices when sending the traffic via the pool. 
By complementing the \nic pool with a shared memory pool, 
\sysname minimizes the across-rack communication period, and the computing nodes can
%
directly access the memory pool with \load/\store instructions (\cxlmem) without moving the data to the local, which unleashes the hardware computing throughput. 

Besides,  
there are admittedly several other critical challenges in realizing \sysname, 
for example, \cxlmem \load/\store is synchronous, and cacheline-based instructions, which is inefficient for accessing a large memory area compared to \dma, 
how to seamlessly use pass-by-reference semantic provided by interconnects for intra-rack communication, given the applications are developed using the socket programming model, \etc (more in \S\ref{sec:motivation}).
We address these challenges in design and implemented a full functional \sysname prototype with four customized MPSoC FPGAs and one server connected by optical fibers, 
instead of using simulator and NUMA nodes in prior works~\cite{ASPLOS24CCNIC, arXiv23case, HPCA24Salus, ASPLOS23TPP}, which enabling to run different applications transparently on \sysname.
We implement \cxl 3.0-like memory protocols, such as \cxlmem and \cxlio, atop an academia lightweight
conceptual hardware protocol stack, which are needed by memory pool and \nic pool. 
To evaluate \sysname, we run both microbenchmarks and real data-intensive applications, 
such as \dnn training, graph processing, and show that \sysname reduces a geometric mean of 30.6\% communication time compared to running them with a conventional \tor-based rack.
\sysname also achieves 40.5\% lower p99 tail latency running Redis.

\section{Background and Motivations}
\label{sec:motivation}

\begin{figure}[t]
\centering
  \subfigure[]{\includegraphics[width=0.22\textwidth]{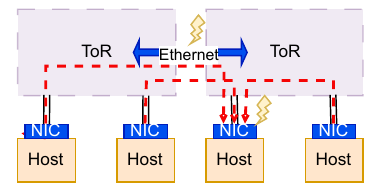}
  \label{fig:ToR-system}
  }
  \subfigure[]{\includegraphics[width=0.22\textwidth]{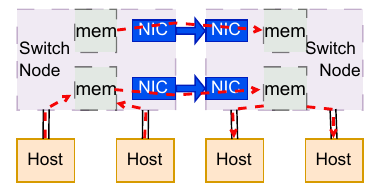}
  \label{fig:DRack-system}
  }
  \hfill
\caption{
Two kinds of rack's architecture proposed by previous work (a) and the \sysname (b).
We abstract the inter-rack network as peer-to-peer connections.
The bottlenecks are tagged with lightning.
}
\label{M_f1}
\end{figure}

In this section, we overview the emerging interconnects, \ie, \cxl and \nvlink, 
and validate the communication inefficiency due to the huge gap in bandwidth between the interconnect link and conventional network link.

\subsection{Emerging Interconnects}
\label{sec:interconnect}
A broad spectrum of applications ranging from traditional high-performance computing (HPC)   
to data-intensive applications running in clouds like machine learning, data analytics, and graph applications have been significantly accelerated by exploiting massive processors and accelerators (\eg, \gpu{}s, TPUs) in parallel. 
To utilize more computing nodes at scale and facilitate efficient communication between processors and accelerators, and among accelerators, 
a new class of fast interconnects is emerging, 
such as Nvidia \nvlink~\cite{Micro17NVLink}, AMD Infinity Fabric~\cite{ISSCC18zeppelin}, and Intel \cxl~\cite{CXLSpec}, 
and integrated into the intra-host interconnect topology to provide unprecedented bandwidth and low latency.

\boldparagraph{Bandwidth and latency.}
To incorporate the evolving computing throughput of accelerators, 
the interconnect bandwidth has seen a substantial improvement, 
In Table~\ref{tab:bandwidth}, 
we show that both \cxl and \nvlink bandwidths exceed CPU memory bandwidth, 
enabling accelerators to access CPU memory at full memory bandwidth.
Besides,
the latency of \cxl and \nvlink exhibit one order of magnitude higher than the memory latency, 
and \cxl incurs an estimated $70$ns for each hop over a switch when scaling \cxl using multi-level switching~\cite{ASPLOS23pond}.
\sysname incorporates these properties of the bandwidth and latency into its design.


\boldparagraph{Resource pooling.}
\cxl fabric introduced in \cxl 3.0 scales up to $4096$ nodes with \cxl fabric~\cite{CXLSpec}.
such as remote memory devices, accelerators (\eg, \gpu{}s, FPGA), 
and hosts, without performance degradation. 
\cxl 3.0 supports memory sharing 
by mapping the memory of nodes into a single physical address space, which can be accessed concurrently by hosts in the same coherency domain.
Compared to \cxl 3.0, 
\nvlink has the following limitations: 
1) its cache coherent extension in fact supports GPUs as \cxl type 2 devices (\cxlcache). 
However, it does not support \cxl type 3 devices that support memory pool; 
2) \nvlink scales to 256 nodes though connects only GPUs as nodes~\cite{Micro17NVLink};
3) \nvlink as a propriety interconnect limits its usage beyond Nvidia’s hardware.
Thus, 
\sysname relies on \cxl fabric as the underlying interconnect due to its generality, scalability and open standard.

\boldparagraph{Pass-by-reference semantics.}
\cn{}s can access any memory location with \cxlmem \load and \store instructions, 
which avoids expensive memory allocations and copying.
\nvlink enables \load/\store primitives between GPUs only, which is similar to \cxlcache instructions. 
\sysname leverages pass-by-reference semantics for intra-rack communications.

\subsection{Communication Bottleneck and Strawman Approach}
\label{sec:bottleneck}
\sysname is motivated by the communication inefficiency 
when running a large-scale compute-intensive workload (\eg, graph, \dnn training) 
in enterprise or public cloud clusters. 
These clusters are commonly built using \tor-based rack architectures, 
as shown in Figure~\ref{fig:ToR-system}, using fast interconnects (\eg, \nvlink) within hosts 
and slow network interconnects (\eg, Ethernet) between hosts.
For example, 
the highest configuration on Tencent Cloud with a 16-host cluster connected with 25\gbps\ Ethernet, 
where each host has 8 Nvidia V100 GPUs connected with NVLink 2.0
that delivers 300 GBps aggregated bandwidth~\cite{arXiv20Tecent}.
Theoretically, when running the job across multiple hosts or racks in parallel, 
its communication efficiency is limited by the slow network link bandwidth.

To validate it, 
we assume the bandwidth ratio between interconnects and network links is $10$. 
We run a Gloo-based ring \allreduce~\cite{gloo} over a dual \tor-based racks architecture (Figure~\ref{fig:ToR-system}) emulated using FPGAs (see \S\ref{sec:implement} and \S\ref{sec:realexp} for the experiment setup).
Figure~\ref{fig:motivation} shows the bandwidth gap is too large to bridge by merely adding one or two \nic, 
while adding too many \nic per host is infeasible due to ``scale tax'' 
such as power consumption, expense and operational costs~\cite{sirius-sigcomm20, rdc-nsdi22, shuffle-infocom24}.

\boldparagraph{Strawman Approach}
By replacing the \tor-based network interconnect with a fast interconnect, 
strawman \sysname interconnects hosts with \cxl fabric
at a large scale (at least 10 host rack)
and reuse their existing \nic{}s to 
form a \nic pool with sufficient large aggregated bandwidth (10-\nic pool) 
across racks.
We built a dual-rack strawman \sysname by using \cxl-\doce(see \S\ref{sec:implement}),
and Figure~\ref{fig:motivation} validates that the resultant completion time of \sysname is approaching the optimal one.

The \nic pool's capacity aggregated by existing \nic{}s may not exceed the bisection bandwidth of the \xu{inter-rack switches}, 
thus contention may happen at the \nic pool if its capacity is saturated by traffic from multiple links.
However, 
communication efficiency is also guaranteed by the fact that ring \allreduce is crafted to minimize the contention at any node, 
which is commonly used in large scale compute-intensive applications~\cite{JPDC09allreduce},
thus only one node at a time (host 1 or host 2) communicates with the \nic pool via a single CXL link.
This also applies to other data parallelism approaches, 
such as hierarchical allreduce, which exploits traffic locality. 
Furthermore, prior research has reported that the utilization of NICs and bandwidth in \tor-based racks is typically low~\cite{SIGCOMM09VL2, IMC09Kandula, SIGCOMM10Benson}, 
a finding that supports the efficiency of the NIC pool.
As a final resort to address contention issues, the system allows for the flexible addition of additional NICs.

\subsection{Challenges}
\label{sec:challege}
Although the high level idea of \sysname is simple, 
there are admittedly several critical challenges to make \sysname work in various environments.

\boldparagraph{C1: Memory bandwidth can be a bottleneck.}
The memory bandwidth may pose a bottleneck
that hinders the full utilization of the \nic pool’s capacity.
\xu{This is because that 1) the interconnect link bandwidth can be larger than the memory access bandwidth, 
and 2) the resultant \nic pool capacity can be larger than the memory bandwidth 
by incorporating concurrent transfers via multiple interconnect links.}
Consequently,
when performing direct memory access (DMA) from all \nic{}s, regardless sending or receiving,
the memory bandwidth becomes the constraining factor, 
reducing the achievable throughput of the \nic pool.
Figure~\ref{fig:motivation} shows that \sysname's performance is degraded when we reducing the memory access bandwidth intentionally.

To address C1, 
we leverage resource pooling in \cxl 3.0 (\S\ref{sec:interconnect}) to 
disaggregate both local and remote memory devices using the CXL fabric. 
These devices are mapped into a unified memory address space, 
creating a shared memory pool accessible in a coherent manner by any \cn{}s. 
Consequently, network traffic incoming from the \nic pool can be distributed across various memory devices, leveraging the aggregated memory bandwidth that surpasses the \nic pool's capacity. 
Likewise, 
\nic pool performs direct memory access (\dma) reads from multiple memory devices to retrieve data for transmission through the pool.

\boldparagraph{C2: \xu{far} memory access exhibits a longer latency.}
%
Table~\ref{tab:bandwidth} shows that 
the memory access with \cxl memory protocols, \ie, \cxlmem, 
demonstrates latency approximately an order of magnitude higher than that of accessing local memory. 
The load and store of \cxlmem are synchronous, cacheline-based instructions with restricted concurrency, \eg, a maximum of 64 instructions~\cite{ISSCC22sapphire}. 
This significantly impairs both intra- and inter-rack communication efficiencies, particularly when data is stored in remote memory (\S\ref{sec:dcache}), 
as shown in Figure~\ref{fig:motivation}, \sysname's performance is degraded to 2.1x.
%
\xu{This is because hosts need to load packets from the memory pool.}
The inefficiency is pronounced for bulk transfer~\cite{footprint}, 
compared to using \dma. 
To address C2, 
we introduce a DRAM cache in \cxl ports, 
enabling caching in \cxlmem to effectively hide the latency (see \S\ref{sec:dcache}).

\boldparagraph{C3: Compability.}
The majority of \dcn applications are developed using the socket programming model, 
which depends on the TCP/IP stack for intra-rack communication (\S\ref{sec:runtime}). 
To address C3 and facilitate the use of pass-by-reference semantics, 
we introduce a kernel module that seamlessly translates socket system calls, such as \send and \recv, into \cxlmem instructions such as \load and \store (see \S\ref{sec:runtime}).

\xu{\boldparagraph{C4: \nic pool bandwidth sharing.}
Cross-rack traffic traverses multiple paths by utilizing all available \nic{}s. 
This raises the risk of out-of-order packet arrivals, 
potentially disrupting the in-order semantics of TCP. 
To address C4, 
\sysname relies on the memory pool to buffer out-of-order arrivals and let the host OS to sequence packets with Multipath TCP (MPTCP) (see \S\ref{sec:inter}).}

\begin{figure}[t]
\centering
\includegraphics[width=0.45\textwidth]{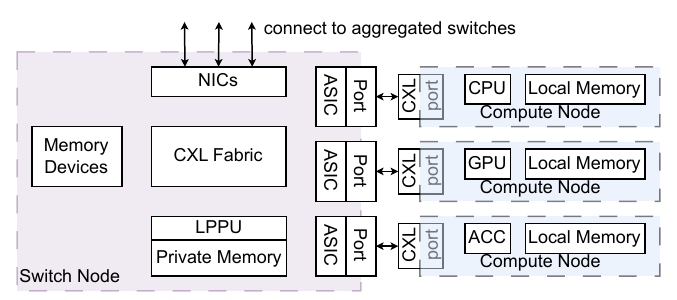}
\caption{An example architecture of the future datacenter. The resources within the \sysname are interconnected with CXL, while the racks are linked through Ethernet.}
\label{fig:DRack_overview}
\end{figure}

\section{\sysname Overview}
\label{sec:overview}

\boldparagraph{Architecture.}
%
Figure~\ref{fig:DRack_overview} delineates the system architecture of \sysname. 
In \sysname's architecture, 
traditional hosts are streamlined to \cn{}s, 
each equipped with a processor or accelerator, such as a CPU or GPU, and integrated local memory.
A segment of this local memory is disaggregated and, in conjunction with additional remote memory devices, constitutes a logical shared memory pool. 
The remaining local memory is designated as the private memory of the \cn. 
This memory pool is integrated into a unified virtual address space, 
enabling consistent access through \cxlmem and \cxlio protocols.
The NIC pool, which can incorporate a variety of NIC types, further enhances the system's flexibility. 
A special \cn, 
low power processor unit (\lppu)~\cite{SIGCOMM23Cowbird}, is dedicated for some bookkeeping tasks such as enumerating, registering, and managing \nic{}s. 
As detailed in \S\ref{sec:nics}, 
the \lppu virtualizes the \nic pool by integrating it into the address space as a singular, ``big'' logical \nic, 
which exposes the pool to all \cn{}s and enables packet-based scheduling for cross-rack traffic, optimizing traffic distribution and bandwidth allocation.
All resource nodes, including \cn{}s, remote memory devices, the \lppu, and \nic{}s, are interconnected by CXL fabric.
\cn{}s access remote resource devices via CXL switching in the fabric, 
forming a logical switch node (\sn).  

\boldparagraph{Workflow.} 
%
\xu{
Although the underlying data structure (lock-free ring buffer) is not new and has been employed before~\cite{LITE17SOSP, FaRM14NSDI, Rambda23HPCA}, our abstraction unifies inter- and intra-rack communication using the same Socket programming model.
}
For intra-rack communication, 
\cn{}s are communicated with pass-by-reference semantics, which uses \cxlmem \load/\store to pass references without moving the data (\S\ref{sec:intra}).
%
For inter-rack communication,
\nic pool \dma transaction is split into two parts (\S\ref{sec:inter}).
Every \cn allocates multiple virtual TX/RX queue pairs in its local memory. 
At the sending rack,
a \cn{} writes the packet descriptor to a virtual TX queue in its local memory.
For each \cn, 
there is a specific ASIC at the port to poll the descriptors from those queues.
When reading a valid descriptor, 
the ASIC writes the descriptor to the working queue of a \nic.
All above reads and writes use \cxlmem \load and \store instructions.
The mapping between \nic{}s and virtual TX/RX queues is determined by \lppu.
%
%
%
\lppu is also responsible for \nic scheduling (\S\ref{sec:nics}) and memory pool allocation (\S\ref{sec:address}).
In \S\ref{sec:inter}, we design a \nic pool scheduling policy on a subflow basis.

At the receiving rack,
receive buffers are allocated from the memory pool 
with the aggregate memory bandwidth larger than the \nic pool capacity.
Thus, 
packets can be \dma{}ed to the memory pool at the full speed of the \nic pool.
There is a dedicated ASIC to poll all the completion queues of the \nic pool.
For every ready descriptor, the ASIC writes the descriptor to the corresponding virtual RX queue of the \cn{} based on the destination IP of the descriptor, and then interrupts the \cn. 

\section{Design Details}
\label{sec:design}


In this section, we present a detailed \sysname designs, especially for those addressing the above challenges.

\begin{figure}[t]
    \centering
    \includegraphics[width=0.45\textwidth]{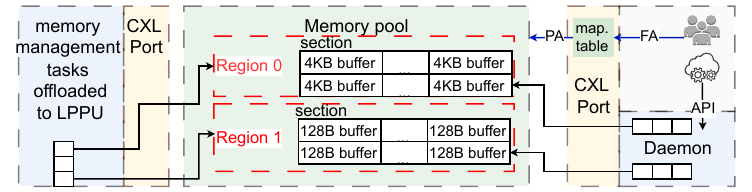}
    \caption{The management machanisms on memory pool}
    \label{fig:unified_memory}
\end{figure}

\subsection{Memory Pool} 
\label{sec:address}

A unified addressed memory pool is the key to realize the pass-by-reference intra-rack communication and 
two-stage inter-rack communication. 
%
The memory pool is mapped to a single fabric address space (FAS) 
and accessed with \cxlmem \load/\store.
We complement \fabric with management mechanisms on the memory pool such as organization, bootstrap, and allocation, which pave the foundation for intra- and inter-rack communications.
%


\boldparagraph{Organization.} 
%
\lppu organizes the shared memory pool as a series of coarse-grained \sect{}s,
each with a size equal to that of a huge page ($2$\MB). 
\lppu also groups $N$ consecutive \sect{}s into a \texttt{region},
where $N$ is configurable, and N=512 in our case.
Upon a \sect{} is allocated to a \cn, 
the \sect can be further divided into \buffer{}s by \cn with the buffer size set based on the running application's demand (with APIs in Table~\ref{d_t1}),
\eg, the \sect with 1\KB\ \buffer.
%

\boldparagraph{Bootstrap.}
\lppu enumerates remote memory devices attached by \fabric and \cn{}s' local memory in the memory pool, and builds a mapping table that maps FAS to memory physical addresses.
Then, every \cn{} enumerates the FAS and private memory,
so that it can access the pool with \cxlmem.
Note that the \cn's private memory has a different address space, 
which is used for kernel, and caches program codes and hot data.
We further use the private memory to implement DRAM cache,
which is used to hide the non-uniformed latency of the memory pool from \cn{}s (\S\ref{sec:dcache}). 

\boldparagraph{Memory allocation.}
%
A daemon running on every \cn is responsible for allocating \sect{}s in advance and managing them.
\lppu updates the mapping from the corresponding FAS to the \sect's physical address in the \cn's mapping table.
Similar to the Linux buddy subsystem, the daemon applies multiple \sect{}s, each having a different \buffer size, the minimum memory management unit, 
\eg, a \sect{} with 2\KB\ \buffer{}s.
An application running in a \cn will consult a daemon for user-level memory allocations 
when it calls \texttt{alloc\_shared\_buffer}.
Specifically,
the daemon will go through the following steps:
1) it chooses \sect{}s with a desirable \buffer size based on the application semantics, 
2) it chooses a \sect{} based on the physical location of the \sect{}s. This is because,
from the \cn's perspective,
the accessing latency profile of the memory pool is different based on the memory's physical locations.
\eg, we prefer local \sect{}s \xu{to store virtual queues} to minimize accessing latency.

\boldparagraph{Discussion.}
\sysname relies on \cxl port to translate the memory request FAS to the physical addresses and verify the permission with the \cn{}'s mapping table, 
\sysname routes requests with the routing algorithm in \fabric~\cite{CXLSpec}, \eg, Port-based.

\begin{figure}[t]
    \centering
    \includegraphics[width=0.45\textwidth]{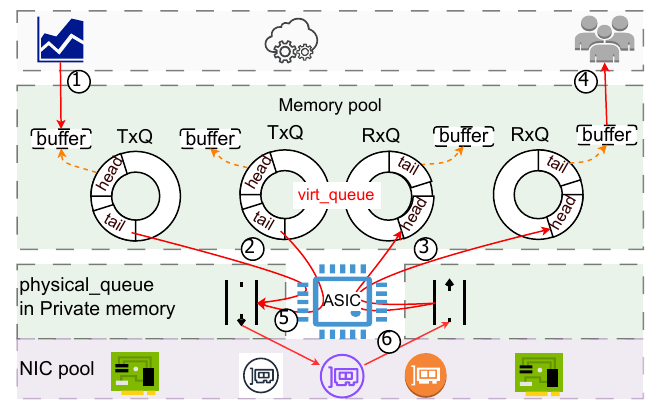}
    \caption{Data structure and metadata used in the interactions with \nic pool and memory pool.}
    \label{fig:nic_pool}
\end{figure}

\subsection{\nic Pool}
\label{sec:nics}
\sysname decouples \nic{}s from hosts to form a \nic pool and interconnects them via \fabric.
\fabric provides \cn{}s with a single \nic abstraction. 
\lppu is responsible for the control plane of the \nic pool 
including configuring \nic{}s and \nic{}s scheduling.
To schedule the \nic pool capacity efficiently and flexibly, 
\lppu schedules \nic{}s on a packet-basis based on their working status, 
\eg, \nic{}'s working queue depth.

The \nic pool and its metadata are shown in Figure~\ref{fig:nic_pool}.
The metadata is used for both intra- and inter-rack communications.
\lppu abstracts \nic pool as a single \nic to each \cn via a set of virtual queues (\texttt{virt\_queue}) that are stored in the memory pool.
\vq includes RX queues (\texttt{RxQ}) and TX queues (\texttt{TxQ}).
\lppu maintains every NIC's working queue (\phyq) in its private memory, 
and uses them for \nic{}s scheduling (see \S\ref{sec:intra}).
\cn{}s and \lppu are responsible for initiating \texttt{virt\_queue}s and \texttt{phy\_queue}s respectively at bootstrap.

\asic is responsible for the data plane between \cn{}s in intra-rack communication (\S\ref{sec:intra}), and \cn{}s and \nic pool in inter-rack communications (\S\ref{sec:inter}).
For example, 
\asic interacts with the \nic{}s by storing and loading descriptors to and from \phyq{}s. 

\subsection{Intra-rack Communication}
\label{sec:intra}

\sysname's intra-rack communication is based on pass-by-reference (pointer) semantics without moving data. 
%
%
To identify the destination of the references, 
every \cn has a unique ID (\eg, IP).
The communication semantic is to transfer the reference addressing the data in the shared memory from \cn $ID_{src}$ to \cn $ID_{dst}$, 
so that \cn $ID_{dst}$ can \load/\store it.
To transfer a reference, 
\sysname stores a transaction descriptor into a \texttt{virt\_queue}, 
where every entry contains a transaction's $ID_{dst}$ and $ID_{src}$, a reference, and the length of the data it references.
As shown in Figure~\ref{fig:nic_pool}
the intra-rack communication can be summarized into the following steps: 

%

\boldparagraph{Data preparation.} 
$ID_{src}$'s application or runtime directly updates the \buffer content via \load/\store to the address \ding{192}. 
\cxl port translates the instruction's FA to physical address,
and encapsulates it in the format of \cxlmem transaction layer packet (TLP).

\boldparagraph{\texttt{TxQ} operations.}
$ID_{src}$ \store{}s the \buffer{}'s descriptor to the head entry of its \texttt{TxQ},
where \txq is implemented as a circular buffer.
The \asic of $ID_{src}$ polls all \txq{}s and obtains a new descriptor by checking the tail of a \texttt{TxQ} \ding{193}. 

\boldparagraph{Split-transaction in \asic.}
The \asic checks whether the descriptor's destination is the \nic pool or another \cn.
For the former one, 
inter-rack communication is instantiated (see \S\ref{sec:inter}).
Otherwise, 
\asic \store{}s the \buffer's descriptor to the head entry of \rxq of $ID_{dst}$ \ding{194} and writes to its interrupt register.

\boldparagraph{\texttt{RxQ} operations.}
$ID_{dst}$'s runtime \load{}s the tail entry of \texttt{RxQ}, and informs the application with the data reference \ding{195}.

\boldparagraph{Data buffer deallocation.} 
Once the application or runtime  no longer uses \buffer, 
it will free the \buffer's address to the daemon.
If there are enough \buffer{}s, 
the daemon will further free the \buffer to \lppu.

The proposed pass-by-reference mechanism leads to zero data copy, and works transparently with the application using the socket programming model (see \S\ref{sec:runtime}).
We can also apply the pass-by-reference mechanism to DPDK~\cite{dpdk} \xu{and RDMA verb} similarly. 

\subsection{Inter-rack communication}
\label{sec:inter}

Despite \nic pool allows a \cn to leverage multiple \nic{}s to achieve a higher throughput.
we face two challenges.
1) As the working status of every \nic's \phyq and the path condition behind it vary over time, 
it is important to design a flexible and efficient \textit{\nic{}s scheduling policy} that can operate on a finer granularity, \eg, packet basis, while fully utilizing the \nic pool's capacity.
2) When a flow is scheduled with multiple \nic{}s, 
it is hard to ensure packets arrive in-order as the path delay can differ.
If the underlying transport enforces in-order arrivals, \eg, TCP, out-of-order arrivals will trigger the \xu{congestion control} to reduce the sending throughput unnecessarily.
Thus, it is also important for the \nic{}s scheduling policy to \textit{minimize the out-of-order arrivals}.
%


\boldparagraph{\sysname's \nic{}s scheduling.}
%
%
\cn{s} equally divide every flow into subflows and map each to a \txq/\rxq pair.
\lppu many-to-one maps \txq{}s to \nic{}s based on the utilization of \nic{}s, 
\eg, the \nic's \phyq depth
(\xu{such as using hardware counters of NVIDIA RNICs through network adapter management tool NEO-Host~\cite{NVIDIANEO}}).
Note that we assume that there exists no link down in the core network, and 
every sub-flow is mapped to a fixed path via ECMP when it traverses the network core.
Thus, 
\sysname fully exploits the \nic pool only if \cn{}s can generate a sufficient number of subflows, 
while ensuring in-order arrivals at the destination within a subflow.
To this end, 
we exploit MPTCP-like indirection in the sending \cn's OS, which can open a sufficient number of TCP subflows,
while resequencing TCP subflows into an in-order data stream at the receiving \cn~\cite{mptcp_rfc}.

For inter-rack communication, 
we assume every \nic has a three \phyq{}s; \texttt{TX\_phq}, a \texttt{RX\_phq}, 
and a \cq, as shown in Figure~\ref{fig:nic_pool}.

\boldparagraph{Packet transmission.}
The \asic of $ID_{src}$ polls all its \txq{}s in the memory pool.
For every \txq{}, 
\lppu has scheduled a \nic to send its \buffer{}s referenced by the entries in the \txq. 
The \asic \store{}s \dma descriptors in that \nic's \texttt{TX\_phq} \ding{196}. 
The \nic's DMA engine is then notified to collect data from the \buffer{}s referenecd and perform network packetizing.

\boldparagraph{Packet receiving.}
\lppu allocates multiple \buffer{}s in the memory pool 
as pool's receiving buffers in advance.
The receiving buffers for \nic{}s should have sufficient memory bandwidth 
than the throughput of the \nic pool. 
\lppu also fills the \texttt{RX\_phq} of every NIC with DMA descriptors referencing receiving \buffer{}s in advance.
For each packet's arrival, 
the \nic's DMA engine directly writes the payload to the \buffer specified by the valid in-queue DMA descriptor. 
A completion notification is stored in the \nic{}'s \cq
as soon as the DMA engine finishes a \dma operation \ding{197}. 
A dedicated \asic is responsible for polling these \cq{}s and informing the \cn{}s of data arrival, 
by storing the \cn's interrupt register based on the packet's IP and storing the references of the \buffer{}s in the \rxq with \cxl stores \ding{194}. 
Eventually, the \cn can load/store \buffer{}s during its computation.


\begin{table}[t]
\caption{memory pool related APIs}
\def\arraystretch{1}\tabcolsep 6pt
\begin{center}
\small
\begin{tabular}{|c|c|c|}
\hline
\textbf{Name}&\textbf{Input} &\textbf{Output}\\
\hline
build\_shared\_skb & 
\makecell[c]{buffer reference,\\length} & \texttt{sk\_buf} reference \\
\hline
kfree\_shared\_skb & 
\texttt{sk\_buf} reference & Void \\
\hline
alloc\_shared\_buffer & size, location & buffer reference\\
\hline
\end{tabular}
\end{center}
\label{d_t1}
\end{table}

\subsection{Software Runtime}
\label{sec:runtime}

To enable applications to enjoy pass-by-reference semantics transparently, 
\sysname provides 
a set of APIs for TCP/IP stack, 
and a driver below the stack to orchestrate intra-rack communication.

The TCP/IP stack operates on the socket buffer object (\texttt{sk\_buf}), 
which includes the \texttt{data} field that stores the reference to the actual data, and 
the data can be located at any memory device within the memory pool.
For performance consideration, 
the TCP/IP stack uses the write-around cache policy for these buffers so that the data is directly written/updated in memory without bringing them to the cache first.
Recall, 
the daemon in every \cn is responsible for applying \sect{}s from \lppu.
Applications and runtimes can apply \buffer{}s from the daemon, 
which can be used for user memory and kernel memory, respectively.
%

\boldparagraph{Socket buffer APIs.}
To enable TCP/IP stack to read these \buffer{}s in the shared memory pool, 
we define three APIs in Table~\ref{d_t1}.
Recall, \cn will be interrupted to read \rxq{}s whenever new references are available.
\cn $ID_{dst}$ calls \texttt{build\_shared\_skb} to wrap the received references with \texttt{sk\_buf} 
before passing it up to TCP/IP stack.
Once TCP/IP no longer uses the buffer, 
it calls \texttt{kfree\_shared\_skb} to hand over the buffer to the daemon.
\cn $ID_{src}$ should call \texttt{alloc\_shared\_buffer} to apply for a buffer from the daemon before populating the buffer with the application's data and TCP/IP headers.
As the memory pool presents a non-uniform accessing latency,
$ID_{src}$ is encouraged to pass the preference on \buffer{} locations based on the application semantics.
For example, a \buffer located at the SN is preferred if several \cn{}s will own the buffer alternately.


\boldparagraph{\sysname driver.}
For configuring \sysname to orchestrate intra-\sysname communications, 
the driver exposes network device operators (\texttt{ndo}) to TCP/IP and spawns the daemon.
The operators include 
1) setting up unified memory address space at the bootup stage 
by registering contiguous FAS to CNs and local shared memory to the memory pool;
2) initiating intra-rack communication transactions once TCP/IP has prepared \texttt{sk\_buf};
3) handling interrupts when receiving communication transactions.

\begin{figure}[t]
    \centering
    \includegraphics[width=0.45\textwidth]{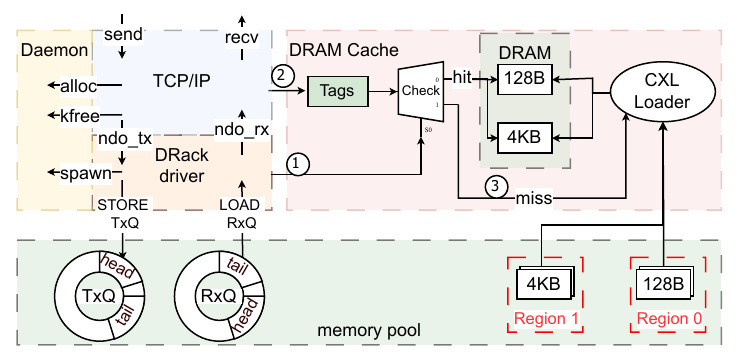}
    \caption{The architecture and working flow of the DRAM cache. }
    \label{d_f2}
\end{figure}

\subsection{CXL-attached DRAM Cache.}
\label{sec:dcache}
\cxl load/store is synchronous and fine granularity memory accessing, 
which limits the throughput of accessing \buffer{}s allocated in remote memory devices, which exhibits locality on memory accessing pattern.
For example, 
copying consecutive data from the kernel to the user mode may exhibit a high spatial-locality.
As \cxl and the CPU structure (MSHR) limit the maximum outstanding number of \load/\store instructions, 
reducing memory latency via caching or migration is the only way to improve the bandwidth. 

Recent works~\cite{MACS23DaeMon, ASPLOS21Kona, ICCD23Morpheus} validated that CXL-attached accelerators can track memory accesses at the cache-line granularity, 
and we can upgrade or replace them without modifying the entire chassis.
Compared to caching, 
page migration~\cite{ASPLOS23TPP,ASPLOS23pond, TACO24Rcmp} requires the runtime to identify hot pages, migrate them to local memory, and update the page table.
we design to install a DRAM cache card in one of the \cxl ports at all \cn{}s.

By analyzing the TCP/IP stack, we summarize the three key points to guide the DRAM cache design. 
\textbf{1) Variable time interval between accessing headers and payloads.}
The runtime processes packet headers and data during the bottom half of the interrupt, 
and the \texttt{recv} syscall, respectively.
Therefore, there will be a time interval between the two, 
in which the data can be evicted out of the cache.
\textbf{2) Buffer release function is explicitly called.} 
As runtime calls the buffer management APIs explicitly,
it will flush the \buffer out of the cache once it calls the buffer release function, 
\ie, \texttt{kfree\_shared\_skb}.
\textbf{3) Variable packet granularity.} 
The packet size may be a few cache lines, 
such as the TCP SYN and ACK.
Caching data with coarse granularity leads to a low cache utilization rate and prolongs the latency, 
thus the packets should be filtered before caching.

As shown in Figure~\ref{d_f2},
we use a multi-way DRAM cache to tackle the potential eviction caused by conflict during variable intervals.
The metadata (\texttt{Tags}) is stored separately in the on-chip memory, 
so that one read can get the whole set's metadata.
The usage of DRAM cache can be summarized as follows.
\ding{192}
The driver configures the FAS of a \texttt{region} -- a consecutive number of \sect{}s, 
whose \buffer{}s will be cached in DRAM cache, and the caching granularity, 
\ie, \buffer size.
For the first access to a \buffer in the \region \ding{193}, 
a miss is triggered and the \buffer is fetched from the memory pool via CXL.io \ding{194}. 
Then, the DRAM cache fills valid metadata and evicts the victim \buffer.
The following accesses to the \buffer \ding{193} will hit the local DRAM cache.
When the runtime decides to free \buffer{}s, \texttt{kfree\_shared\_skb} will be explicitly called and flush the DRAM cache.

Currently, 
\sysname uses the DRAM cache to improve the throughput of loading/storing a large memory segment.
However, the cache design is general and independent from the runtime,
thus applications can transparently use the DRAM cache by configuring their FAS of \region{}s. 

\xu{

\begin{figure}[t]
    \centering
    \includegraphics[width=0.45\textwidth]{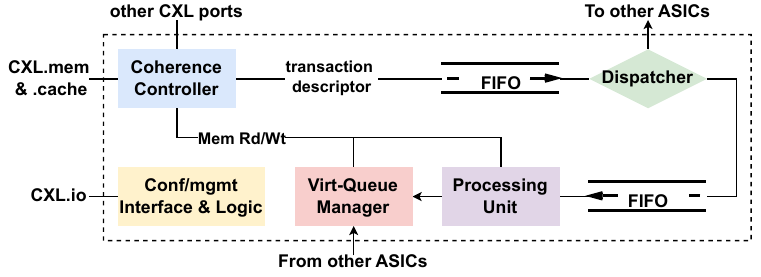}
    \caption{The architecture of each port's data plane accelerator. }
    \label{fig:asic}
\end{figure}

\subsection{ASIC Architecture}

Each CXL port of \sn has a data plane accelerator implemented using ASIC or FPGA (Figure~\ref{fig:asic}), 
which is responsible for the intra- and inter-rack communication defined in Section~\ref{sec:inter} and \ref{sec:intra}.
Despite functions defined by CXL Switch, such as forwarding requests among ports, address decoding, etc,
the Coherence Controller handles the memory requests from other components and forwards the polled transaction descriptors to the Dispatcher.
the Coherence Controller may instantiate an address translation cache to support applications' virtual address~\cite{ATS, Kong23NSDI}.
The Dispatcher forwards descriptors to other \asic{s} or the Processing Unit based on which type of communication the descriptors define.
The Processing Unit should be general-purpose 
so that it can translate transaction descriptors to specific descriptors of various \nic{s}.
Besides, 
the Processing Unit polls or read/write \nic{s'} \phyq via the Coherence Controller, 
and translates completion notifications to transaction descriptors.
\sn may implement multiple Processing Units distributed within each \asic, 
or a giant Processing Unit shared by all \asic{s}.
\sn{'s} private memory can be a separate CXL memory device or RAM integrated with the Processing Unit.
The Virt-Queue Manager contains the context of each virtual queue, including the head, tail pointers, and queue depth.
It needs to poll the \txq{s} and store received transaction descriptors from other \asic{s} and the Processing Unit in the \rxq{s} via the Coherence Controller.
}

\section{Discussion}
\label{sec:discussion}
Separating data and control plane in the SN is a common optimization to assign tasks to the appropriate computing resources~\cite{Guo22Clio}.
\sysname assigns the \lppu with the control plane, which is not often evoked, complicated, full of control paths, and hard to accelerate.
\sysname leaves the data plane to the dedicated ASIC, which provides high data processing throughput.
However, concerning the polling tasks, the multi-core CPU or smartNIC may have enough capability~\cite{Chen23Cowbird}.
For example, we can bind one core to poll one CN or NIC.

In \sysname, the sub-flow is mapped to one NIC based on its physical\_queue depth. 
However, the factors, such as congestion and hops in the core network, should also be considered.
The commercial data centers' topologies are required to simulate various congestion situations.
We hypothesize no congestion and one hop in this paper and leave them for future work.

\section{Related Work}
\label{sec:related}
Prior works leverage multi-port NICs or add additional \nic{}s to expand host egress capacity~\cite{NSDI15GRIN}, which is effective given the huge gap between interconnects and networks.
GPU clusters such as ELUPS~\cite{Micro14Zang} and EFLOPS~\cite{HPCA20EFLOPS} expand the rack-level bandwidth by bounding each GPU to a single \nic to avoid PCIe contention.
Disaggregated Rack Architecture (\dra)~\cite{Micro14Zang, ISCA13Ladon, ANCS14Marlin} uses \pcie interconnect to allow a host to use multiple \nic{}s of the rack, which is configured in advance. 
Similar to \tor-based racks, 
the host in \dra can be limited by \xu{Integrated I/O controller.}

\section{Conclusion}
\label{sec:conclude}

We present a novel interconnect architecture, \sysname, 
that bridges the bandwidth gap between interconnects and networks with a \nic pool, and optimize its efficiency with several novel designs such as a shared memory pool and DRAM Cache. 
We build a dual-rack prototype and validate its performance with both microbenchmarks and real experiments.

\bibliographystyle{plain}
\bibliography{local,main}

\end{document}